\documentclass[prd,aps,showpacs,notitlepage,superscriptaddress,nofootinbib]{revtex4-1}
\usepackage{epsfig}
\usepackage{color}
\usepackage{url}
\usepackage{amsmath}
\usepackage{multirow}
\def\bea#1\eea{\begin{align}#1\end{align}}

\newcommand{\bef}{\begin{figure}[htb]\centering}
\newcommand{\eef}{\end{figure}}

\begin{document}
\title{Vector boson-tagged jet production in heavy ion collisions at the LHC}

\date{\today}

\author{Zhong-Bo Kang}
\email{zkang@physics.ucla.edu}
\affiliation{Department of Physics and Astronomy, University of California, Los Angeles, California 90095, USA}            
\affiliation{Mani L. Bhaumik Institute for Theoretical Physics, University of California, Los Angeles, California 90095, USA}
\affiliation{Theoretical Division, Los Alamos National Laboratory, Los Alamos, New Mexico 87545, USA}    

\author{Ivan Vitev}
\email{ivitev@lanl.gov}                   
\affiliation{Theoretical Division, Los Alamos National Laboratory, Los Alamos, New Mexico 87545, USA}       
                   
\author{Hongxi Xing}
\email{hxing@northwestern.edu}
\affiliation{Department of Physics and Astronomy, Northwestern University, Evanston, Illinois 60208, USA} 
\affiliation{High Energy Physics Division, Argonne National Laboratory, Argonne, Illinois 60439, USA}

\begin{abstract}
Vector boson-tagged jet production in collisions of heavy nuclei opens new opportunities to study parton shower formation and  propagation in strongly interacting matter. It has been argued to provide a golden channel that can constrain  the energy loss of jets in the quark-gluon plasma created in heavy ion reactions.  We present theoretical results for isolated photon-tagged and $Z^0$ boson-tagged jet production in Pb+Pb collisions with $\sqrt{s_{NN}} = 5.02$ TeV at the LHC. Specifically, we evaluate the transverse momentum imbalance ${\rm x_{JV}}$ distribution and nuclear modification factor  ${\rm I_{AA}}$ of tagged jets and compare our theoretical calculations to recent experimental measurements by ATLAS and CMS collaborations.  Our analysis, which includes  both collisional and radiative energy losses, sheds light on their relative importance versus the strength of jet-medium interactions and helps quantify the amount of out-of-cone radiation of  predominantly prompt quark-initiated jets.    
\end{abstract}

\maketitle

\section{Introduction}
The production of a vector boson (either a photon $\gamma$ or an electroweak boson  such as the $Z^0$) in association with a jet has been extensively studied in proton-proton collisions at the Large Hadron Collider (LHC), by both the ATLAS~\cite{Aad:2013ysa,Aad:2013zba,Aaboud:2016sdm} and CMS~\cite{Chatrchyan:2013mwa,Chatrchyan:2013oda,Khachatryan:2014zya,Khachatryan:2015ira} collaborations. Such $\gamma$+jet and $Z^0$+jet processes are among the most powerful channels that can be used to test the fundamental properties of Quantum Chromodynamics (QCD). They also serve as crucial inputs for the precise determination of 
the parton densities in the proton, and  can help improve  the constraints on the gluon distribution function. It is, thus, not surprising that significant theoretical effort has been invested in precisely computing the differential cross sections for these processes~\cite{Catani:2002ny,Boughezal:2015ded,Ridder:2015dxa}. 

Vector boson-tagged jets are also particularly well suited to studying many-body QCD at high energies in heavy ion collisions, where a deconfined quark-gluon plasma (QGP) is expected to be formed. On one hand,  the tagging bosons escape the region of the hot dense medium unscathed. This has been confirmed through the absence of significant modification of both $\gamma$ and $Z^0$ boson production in Pb+Pb collisions relative to the binary collision-scaled proton-proton (p+p) baseline by both ATLAS and CMS collaborations~\cite{Aad:2012ew,Aad:2015lcb,Chatrchyan:2014csa}.  On the other hand, the parton shower that recoils opposite the vector boson in heavy ion collisions gets modified, or quenched, due to the elastic and inelastic interactions with the QCD medium. Since at leading order the vector boson and the jet are produced back-to-back in the azimuthal plane and have equal transverse momenta in the standard collinear factorization framework,  it was argued more than a decade ago~\cite{Srivastava:2002kg} that a virtual photon 
that decays to dileptons ($\gamma^* \rightarrow  \ell^+\ell^-$) will provide very tight constraints on the energy of the away-side parton shower. Theoretical 
studies of cold nuclear matter effects have shown that they don't significantly affect vector boson-tagged jet  distributions~\cite{Dai:2012am}.    

However,  only recently  have measurements of approximately back-to-back isolated $\gamma$+jet and/or $Z^0$+jet~\footnote{The so-called fragmentation contribution to $Z^0$-boson production is generally small even at the LHC energies~\cite{Berger:2015nsa}.} final states, considered ``golden channels'' for the study of jet quenching and the extraction of the properties of the hot dense medium,  become possible.   It was also realized that higher order processes  will alter the perfect transverse momentum balance $p_T^J = p_T^V $ and lead to a distribution of recoiling jets~\cite{Neufeld:2010fj}.  A useful feature of this distribution for the  purpose of our study is that it is narrowly peaked and the shift of the peak will contain detailed information about jet energy loss.   Furthermore, jets produced 
opposite to the isolated $\gamma$ or $Z^0$ bosons are much more  likely to originate from quarks, while dijets usually involve significant quark and gluon fractions that vary strongly with transverse momentum. In this regard, vector boson-tagged jets can help constrain the flavor dependence of the jet quenching mechanism. Previous studies of vector boson tagged jet production in heavy ion collisions have been carried out based on a perturbative QCD framework~\cite{Dai:2012am,Neufeld:2012df}, a  Boltzmann transport model~\cite{Wang:2013cia}, an event generator JEWEL~\cite{KunnawalkamElayavalli:2016ttl}, and a hybrid strong/weak coupling model~\cite{Casalderrey-Solana:2015vaa}. Photon-tagged heavy flavor jets have also been proposed as ways to increase the fraction of prompt $b$ quarks~\cite{Huang:2015mva}. Last but not least, the substructure of $\gamma$-tagged jets was found to be more sensitive to large angle radiation in comparison to inclusive jets~\cite{Chien:2015hda}.

Isolated $\gamma$-tagged and $Z^0$-tagged jets in Pb+Pb collisions at the center-of-mass energy per nucleon pair $\sqrt{s_{NN}}=5.02$ TeV have been recently measured at the LHC by the ATLAS and CMS collaborations~\cite{Sirunyan:2017jic,CMS:2016ynj,ATLAS:2016tor}. Motivated by these new measurements, in this paper, we provide our theoretical calculations and comparison to the experimental data. In particular, by including both collisional and radiative energy loss effects, we evaluate the so-called transverse momentum imbalance ${\rm x_{JV}}$ distribution in both p+p and Pb+Pb collisions, where ${\rm x_{JV}} = p_T^J/p_T^V$ with $p_T^J$ and $p_T^V$ the transverse momentum of the jet and the vector boson, respectively. We also calculate the nuclear modification factor ${\rm I_{AA}}$ and compare to the experimental findings. Within the theoretical model calculation we present our results for the
transverse momentum imbalance shift    $ \Delta \langle  {\rm x_{JV}} \rangle$ and the relative contribution of radiative and collisional
energy losses of typical energy jets.
 
The rest of our paper is organized as follows. In Sec.~\ref{pp}, we present the evaluation of the differential cross sections for isolated $\gamma$-tagged and $Z^0$-tagged jet production in p+p collisions using Pythia 8~\cite{Sjostrand:2007gs} and determine the flavor origin of the recoil jet production for the proper implementation of the energy loss effects. In Sec.~\ref{AA}, we provide information on how we implement the medium effects to obtain the modification of vector boson tagged jet production in dense QCD matter. In Sec.~\ref{numerical}, we present our phenomenological results and give detailed comparison with the most recent experimental measurements for the isolated $\gamma$ and $Z^0$ boson tagged jet production in heavy ion collisions at the LHC. We summarize our paper in Sec.~\ref{summary}.

\section{Isolated photon-tagged and $Z^0$-tagged jet production in p+p collisions}
\label{pp}

In this section we present the evaluation of the differential cross sections for isolated photon-tagged and $Z^0$-tagged jet production 
in p+p collisions using Pythia 8~\cite{Sjostrand:2007gs}. Pythia 8 is a widely-used  high energy phenomenology event generator, which can describe well the main properties of the event structure. This event generator utilizes leading-order perturbative QCD matrix elements+parton shower, combined with the Lund string model for hadronization. The simulations presented in this paper are performed with the CTEQ6L1 parton distribution functions~\cite{Pumplin:2002vw} and with the anti-$k_T$ jet clustering  algorithm~\cite{Cacciari:2008gp}. In the p+p baseline simulations, we select the vector boson (isolated-photon and $Z^0$-boson) and jet according to the desired kinematics to match the experimental measurements, and we have simulated around $10^7$ events for both isolated photon-tagged and $Z^0$-tagged jets to reduce the statistical uncertainties/fluctuations. 

Measurements of vector boson-tagged jet production in p+p collisions at different center-of-mass energies have been carried out at both the Tevatron and the LHC. We present in Fig.~\ref{fig:pp-xsec} the comparisons to CMS measurements~\cite{Khachatryan:2014zya, Khachatryan:2015ira} to show the validation of Pythia simulation against  experimental data. The left panel in Fig.~\ref{fig:pp-xsec} is the differential cross section $d\sigma/dp_T^{J}$ as a function of leading jet transverse momentum $p_T^{J}$ for $Z^0$+jet production in p+p collisions at the LHC at $\sqrt{s}=7$~TeV. The right panel  corresponds to the differential cross section $d\sigma/dp_T^{\gamma}$ as a function of isolated-photon transverse momentum $p_T^{\gamma}$ for $\gamma$+jet production in p+p collisions at the LHC at $\sqrt{s}=8$ TeV. In our simulations, the specific kinematic requirements are implemented to match the experimental measurements in selecting  V+jet events.  For details on the kinematic cuts, see Ref.~\cite{Khachatryan:2014zya} for $Z^0$+jet and 
Ref.~\cite{Khachatryan:2015ira} for $\gamma$+jet production. As can be seen in Fig. \ref{fig:pp-xsec}, the Pythia 8 event generator gives reasonably good description of the CMS experimental data. 

\begin{figure}[!t]
\psfig{file=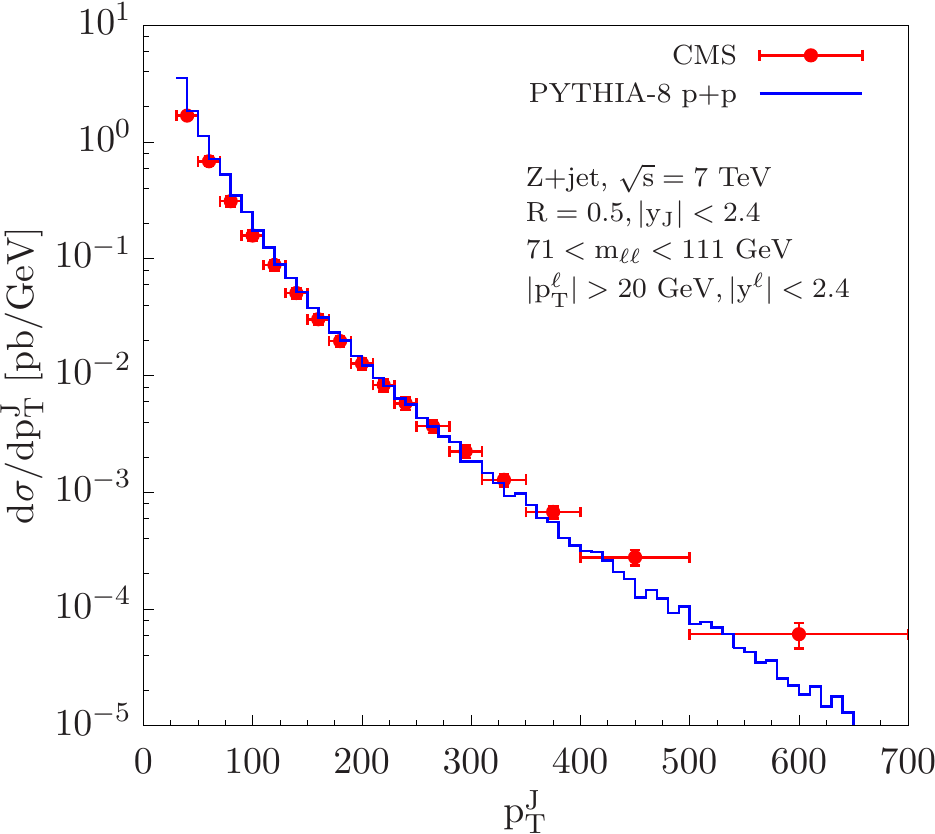, width=2.5in}
\hskip 0.2in
\psfig{file=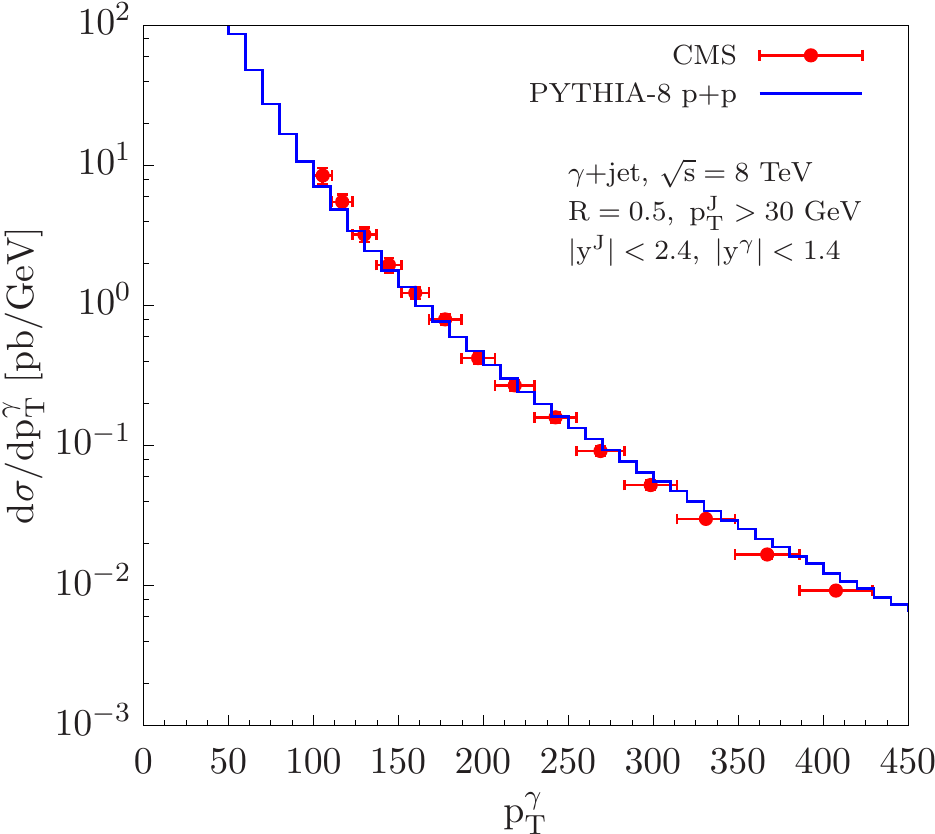, width=2.5in}
\caption{Comparison between Pythia 8 simulations and CMS measurements of V+jet production in p+p collisions at the LHC. Left: the 
$Z^0$+jet differential cross section at $\sqrt{s} = 7$ TeV as a function of $p_T^{J}$. Right: the isolated photon+jet differential cross section at $\sqrt{s} = 8$ TeV as a function of $p_T^{\gamma}$. The blue curves are from Pythia 8 simulations, the red data points are from the CMS collaboration~\cite{Khachatryan:2014zya, Khachatryan:2015ira}.}
\label{fig:pp-xsec}
\eef

Before the implementation of energy loss effects through the medium-induced parton shower on vector boson-tagged jet production in Pb+Pb collisions at the LHC, we need the detailed baseline information for $Z^0$+jet and $\gamma$+jet production in p+p collisions for different partonic subprocesses. In our simulations, specific kinematic cuts for the $Z^0$+jet and $\gamma+$jet event selections are applied as in Refs. \cite{Sirunyan:2017jic} and \cite{CMS:2016ynj}, respectively. 
In particular, a minimum separation of the azimuthal angle between the vector boson and the jet, ${\rm \Delta \phi_{JV}>7/8\pi}$, is required to select back-to-back V+jet events. In each event of the $Z^0$+jet simulation, the $Z^0$-boson is required to have: the invariant mass of the decayed dileptons  $70< m_{\ell\ell}<110$ GeV, $p_T^{e}>20$ GeV, $p_T^{\mu}>10$ GeV, $|y^{e}|<2.5$, $|y^{\mu}|<2.4$, and the transverse momentum of the $Z^0$ boson $p_T^{Z}>60$ GeV; the recoil jet is reconstructed using the anti-$k_T$ algorithm with a jet radius parameter $R=0.3$, $p_T^{J}>30$ GeV and $|y|^{J}<1.6$ in the same event. For $\gamma$+jet production, the photon is required to have $|y^{\gamma}|<1.44$. To minimize the fragmentation contribution to the photon, an isolation cut is applied where the sum of the transverse momenta of the generated particles in a cone of radius $\Delta R=0.4$ around the photon is required to be less than 5 GeV. Unless explicitly specified, these kinematical cuts apply to all the results shown in the rest of the paper.

In V+jet production, there are two dominant  channels at leading order that are implemented in Pythia, i.e. $q+\bar q \to V+ g$ and $q(\bar q)+g \to V+q(\bar q)$. We have checked that the $g+g\to V+g$ channel contributes to the cross section only marginally and, thus, can be safely neglected. As can be seen in Fig.~\ref{fig:pp-frac} (left), the cross section of $Z^0$+jet production is dominated by $q(\bar q)+g \to Z+q(\bar q)$ channel (around 80\%) for a wide $p_T$ range. In other words, the produced jet  predominantly originates from a light quark. The fraction for $\gamma+$jet production behaves similarly to the case of $Z^0$+jet production, with even higher fractions from the $q(\bar q)+g \to \gamma+q(\bar q)$ channel. This implies that in heavy ion collisions at LHC energies, the medium modification of V+jet production is dominated by quark energy loss. We will present the detailed discussions about the medium effects on V+jet production in the next section.

\begin{figure}[!t]
\psfig{file=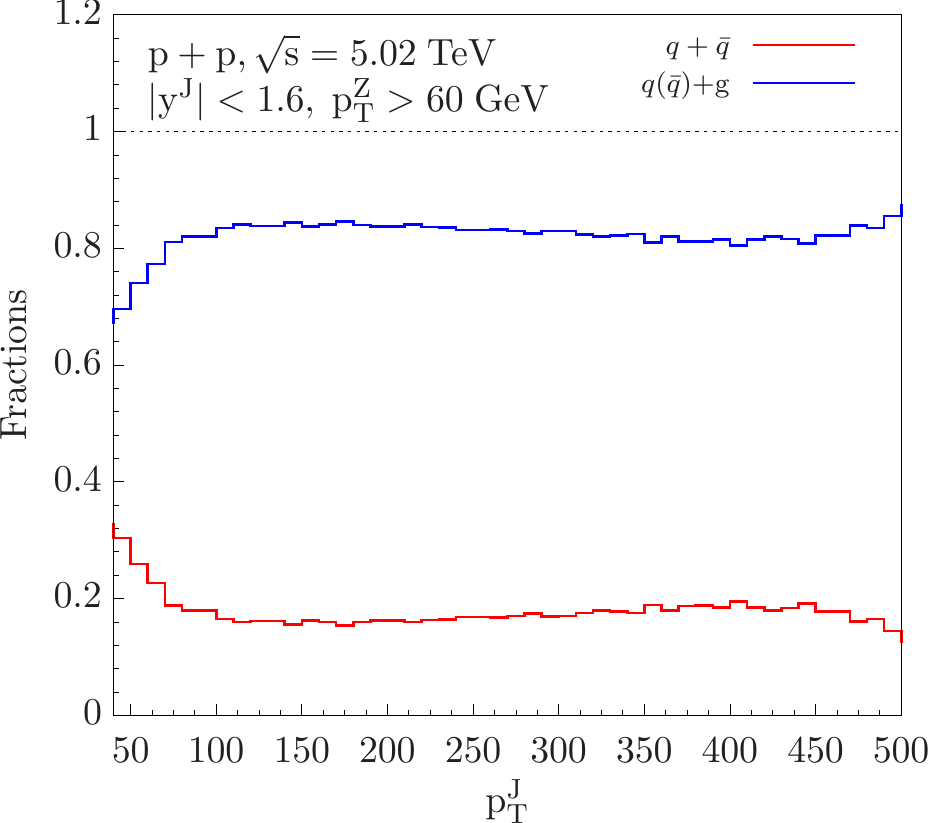, width=2.5in}
\hskip 0.2in
\psfig{file=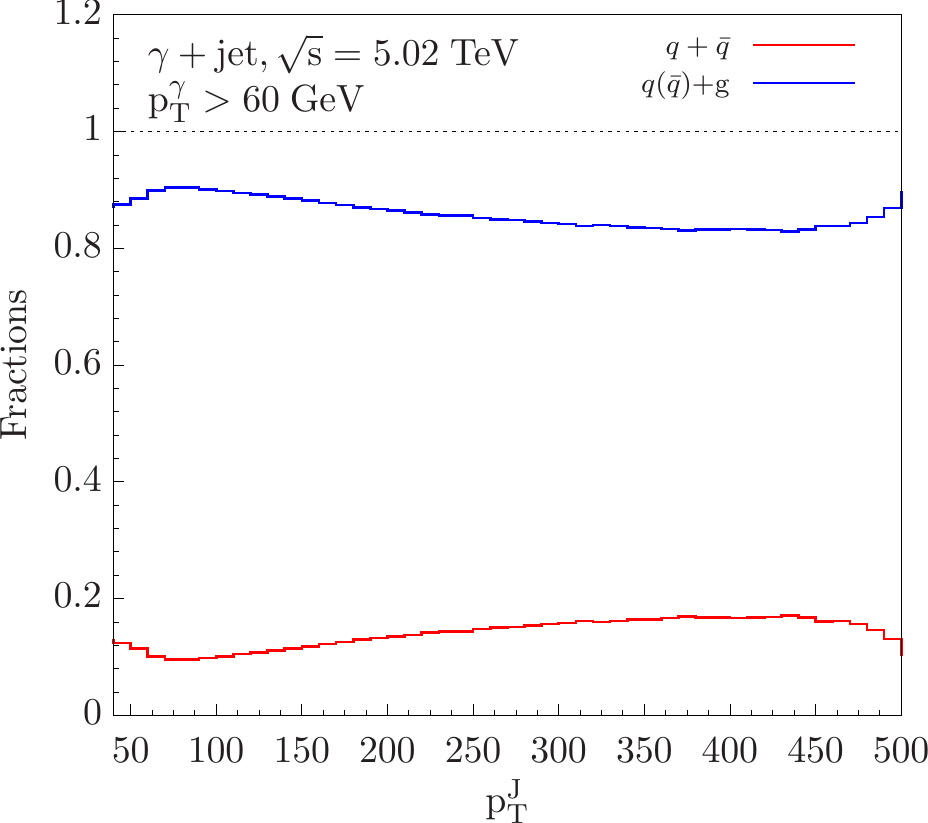, width=2.5in}
\caption{The fractional contributions of different subprocesses to the $Z^0$+jet (left) and isolated-$\gamma$+jet (right) production cross sections in p+p collisions at $\sqrt{s}=5.02$ TeV. Kinematical cuts are implemented in our simulations as in CMS measurements, see Ref.~\cite{CMS:2016viv} for $Z^0$+jet and Ref.~\cite{CMS:2016ynj} for isolated-$\gamma$+jet.} 
\label{fig:pp-frac}
\eef

\section{Modification of tagged jet production in dense QCD matter}
\label{AA}

In the presence of dense QCD matter, such as the QGP created in heavy-ion collisions,
the vacuum parton shower is modified. Early investigations focused on 
non-Abelian energy loss processes~\cite{Zakharov:1997uu,Baier:1996kr,Gyulassy:2000fs,Wiedemann:2000za,Wang:2001ifa,Arnold:2002ja}. 
The soft gluon emission limit was subsequently  relaxed to  allow for a unified description of parton branching 
processes~\cite{Ovanesyan:2011kn,Ovanesyan:2015dop}.  In addition to radiative processes, collisional energy loss 
has also attracted a lot of attention~\cite{Braaten:1991we,Wicks:2005gt,Adil:2006ei,Thoma:2008my,Berrehrah:2013mua,Neufeld:2014yaa}  and was found to play a more significant role for lower parton energies.

At present, the application of full in-medium splitting functions~\cite{Chien:2016led,Kang:2017frl} has not been combined 
with collisional energy loss processes. For this reason, we here follow the soft gluon emission radiative energy loss   approximation.  The benefit of 
this approach is that it allows us to consider multiple emissions.  For a given impact parameter $|{\bf b}_\perp |$, taken along 
the $x$-axis in the transverse plane of nucleus-nucleus   collisions, we evaluate the  cross sections as follows  
\begin{eqnarray}
\frac{d\sigma^{AA}(|{\bf b}_\perp |) }{dp_{T}^{V}dp_{T}^{J}} 
&=&  \int d^2 {\bf s_\perp}   T_A\left( {\bf s}_\perp - \frac{{\bf b}_\perp}{2}  \right)
T_A\left( {\bf s}_\perp + \frac{{\bf b}_\perp}{2}  \right)  \sum_{q,g} \int_0^1 d\epsilon \frac{P_{q,g}(\epsilon; s_\perp, |{\bf b}_\perp | )}
 {1-f_{q,g}^{\rm loss}(R;s_\perp, |{\bf b}_\perp | )\, \epsilon}
\nonumber \\ 
 &&   \times \,
\frac{ d\sigma^{NN}_{q,g}  \left( p_{T}^{V}, p_{T}^{J}/ \{1-f_{q,g}^{\rm loss}(R; s_\perp, |{\bf b}_\perp | )\,  \epsilon \} \right) }
{dp_{T}^{J}dp_{T}^{V}} \, ,
\label{eq:modify}
\end{eqnarray}
Let us now discuss Eq.~(\ref{eq:modify}). Hard processes in heavy ion collisions follow a binary collision density 
distribution in the transverse  plane at position ${\bf s}_\perp$. This means that the point-like large $Q^2$ scattering is 
distributed  $   \propto T_A\left( {\bf s}_\perp - {{\bf b}_\perp}/{2}  \right) T_A\left( {\bf s}_\perp + {{\bf b}_\perp}/{2}  \right)  $,
where   $T_A\left(  {\bf s}_\perp  \right) = \int_{-\infty}^{\infty}   \rho_A({\bf s}_\perp,z ) dz$. In our calculation we use
an optical Glauber model and inelastic nucleon-nucleon scattering cross sections  $\sigma_{\rm in} = 70 $~mb to obtain   
average number of binary collisions at $\sqrt{s_{NN}} = 5.02$~TeV. 

In heavy ion collisions a fraction  $\epsilon$  of the energy of the parent parton can be redistributed  through medium-induced 
bremsstrahlung. This process is independent on whether a jet is reconstructed or not, but reflects instead the
parton energy, color charge, path length and medium properties  dependence of the non-Abelian bremsstrahlung.
The probability distribution  $P_{q,g}(\epsilon) $  of this energy fraction satisfies the following properties 
\begin{eqnarray}
  &&\int_0^1 P_{q,g}(\epsilon) d\epsilon =1 \,,   \qquad    \int_0^1 \epsilon P_{q,g}(\epsilon)  d\epsilon = 
 \frac{\langle \Delta E^{\rm rad}_{q,g} \rangle }{E_{q,g}}  \,,
\label{prob} 
\end{eqnarray}
for every jet energy and every transverse position  ${\bf s}_\perp$ at a given impact parameter. To calculate this 
probability, we first need to evaluate the medium-induced gluon radiative spectrum 
\begin{eqnarray}
\frac{dN^g_{q,g}(\omega,r)}{d\omega dr}  &\propto&  C_R  \alpha_s    \int_0^\infty d \Delta z \, \frac{1}{\lambda_g(\Delta z)}      \left[  \int d^2 {\bf q}
  \left( \frac{1}{\sigma_{el}(\Delta z)} \frac{d \sigma_{el}(\Delta z) }{d^2 {\bf q}} - \delta^2 ({\bf q}) \right)    \right]  \;  
   \nonumber  \\  && \hspace*{2cm} \times     
    \frac{ 2 {\bf k}  \cdot  {\bf q} } {  {\bf k}^2 (   {\bf k} - {\bf q} )^2  }   \left\{  1-\cos \left[  \frac{ (   {\bf k} - {\bf q} )^2  }{2\omega} \Delta z   \right]   \right\}
\label{1stOOp}
\end{eqnarray}
of parent quarks and gluons. Here, $\omega$ and $r$ are the energy and angle of the radiated gluon and for small
angles $|{\bf k}| =\omega r$. 
   This calculation is performed to first order in opacity and the integral over $\Delta z$  is along the path of the jet propagation through
   the QGP medium  from the  hard  collision point. In the soft gluon emission limit only the  gluon scattering length 
  $\lambda_g$ plays a role and  quarks
 and gluons lose energy strictly proportional to their  squared color charge.
  The Casimir $C_R$ in  Eq.~(\ref{1stOOp})  is $C_F=4/3$ for parent quarks and   $C_A=3$ for parent gluons.      
  The momentum transfers ${\bf q }$ between the  jet and the medium  are distributed according to a normalized differential 
  elastic scattering cross section, including a unitarizing forward scattering contribution.

The spectrum is first averaged  over the  collision geometry, see for example the first line of Eq.~(\ref{eq:modify}).  In the QGP
we include an effective gluon mass via ${\bf k}^2 \rightarrow  {\bf k}^2  +  \mu_D^2(\Delta z)$.  In this 
evaluation the exact leading power and sub-leading logarithmic  dependence in the path length, density and 
coupling $g$ between the jet and the medium is retained. In the gluon emission vertex  the strong coupling is taken to run with 
the transverse gluon mass.  In  the application of Eq.~(\ref{eq:modify})  the  point-by-point in collision geometry  
radiative gluon spectrum is unfolded to leading power in the path length, coupling $g$ and gluon density, which goes as 
 $ \propto g^4 \int d \Delta z \;  \Delta z \,   \rho_g( {\bf s}_\perp + {\bf n}_\perp \Delta z, \tau_0 + \Delta z )$. Here 
 ${\bf n}$ is the direction of jet propagation and we take the  medium formation time $\tau_0 =0.3$~fm.   On a 
 position-by-position basis  and for every parent parton energy $E$ we can then obtain
 \begin{eqnarray}
 && \frac{dN^g_{q,g}(\omega)}{d\omega} =  \int_0^{R_{\rm max}} dr \,  \frac{dN^g (\omega, r) }{d\omega dr} \;,  \qquad 
 \langle N^g_{q,g} \rangle =  \int_0^E  d \omega \; \frac{dN^g(\omega)}{d\omega} \, .
 \label{gnumber}
 \end{eqnarray}
 In Eq.~(\ref{gnumber})  $R_{\rm max} \gg 1 > R $ is a large radius chosen to capture the parton shower. In our calculation we use
 $ R_{\rm max} = 2 $. 
 In the Poisson approximation the probability density for fractional energy loss $\epsilon=\sum_i \omega_i / E$ 
 can be obtained as follows
\begin{eqnarray} 
&& P_{q,g}(\epsilon)= \sum_{n=0}^\infty P_{q,g}^n(\epsilon)  \, , \quad  P_{q,g}^0(\epsilon)=e^{-\left\langle N_{q,g}^g \right\rangle} \delta(\epsilon)  
 \, , \quad  P_{q,g}^{n+1}(\epsilon) = \frac{1}{n+1} \int_{0}^{E} d\omega  \;  \frac{dN^g_{q,g}(\omega)}{d\omega} \;  
P_{q,g}^n \left(\epsilon-\frac{\omega}{E}\right)  \; .
\end{eqnarray}

For inclusive and tagged hadron production,  unless one focuses on the  $p_T$  region below 5 GeV,  the fragmentation of radiated gluons does not contribute because they are typically soft. Since jets are defined by the amount of energy reconstructed inside the jet cone of radius parameter $R$,
the evaluation of cross sections  with jets in the final state  critically depends on the determination of how much of the energy of the  
medium-induced parton shower actually falls outside  of the jet~\cite{Vitev:2008rz}.  We here denote this fraction by $f_{q,g}^{\rm loss}(R)$, suppressing all other dependencies of this quantity.  Let us first concentrate only on radiative processes.
In this case we have 
\begin{equation}
f_{q,g}^{\rm loss}(R;{\rm rad}) =  \left(  \int_R^{R_{\rm max}  }  dr  \int_0^E  d\omega \,     \frac{dN^g_{q,g}(\omega,r)}{d\omega dr}    \right)  \Bigg/
 \left(  \int_0^{R_{\rm max} }   dr  \int_0^E  d\omega \,     \frac{dN^g_{q,g}(\omega,r)}{d\omega dr}   \right) \;.
\label{frad}
\end{equation}
The radiative out-of-cone energy loss is purely determined by the wide-angle medium-induced radiation pattern.

Collisional interactions  take energy away from the jet through the excitation of the QGP medium and dissipation of the energy
away from the collision axis.  The amplification of the collisional energy loss effects comes from the multiple emitted 
gluons~\cite{Neufeld:2011yh}. In our simulation we assume that all of the energy is taken away from the jet. This is justified because we 
consider jets of small radius $R\ll 1$, whereas Mach cones shockwaves propagate at angles $\theta_M = \arcsin c_s$.   Thus, taking  $c^2_s \approx 1/3$, we find $\theta_M \sim 1$ and the energy deposited by collisional processes is transported out of the jet cone. 
It is important to realize that considering radiative energy loss only and radiative+collisional energy loss of the type discussed here covers the two extreme possible cases. If part of the energy is not fully dissipated, this will be a situation that falls in-between those two scenarios.  
From the average gluon number and the mean total radiative energy loss we can determine the  mean energy per 
emitted gluon $\langle \omega_{q,g} \rangle = \langle \Delta E_{q,g} \rangle /  \langle N^g_{q,g} \rangle   $.    
Parametrically, the collisional energy loss rate  to leading logarithmic accuracy goes as 
$d \Delta E^{\rm coll} / d \Delta z  \propto C_R g^2 \mu_D^2 \ln (E/\mu_D)$.   In Ref.~\cite{Neufeld:2011yh} we set a simulation of
the collisional energy loss of the medium-induced shower as the parent parton propagates through the  medium and showers off
gluons. The average number of gluons, rounded to an integer number,  were distributed along the path of jet propagation at positions $z_i$ and the net 
collisional energy loss obtained. Since the softer medium-induced gluons thermalize first, for later convenience we can express this total collisional energy loss as an integral over the spectrum of the medium induced gluons   
\begin{equation}
\Delta E^{\rm coll}_{q,g}({\rm tot.})   = \sum_{i=1}^{N^{\rm tot. \; partons}_{q,g} \ }  \int_{z_i}^\infty   \frac{d \Delta E_i^{\rm coll}}{ d \Delta z}   d   \Delta z 
\;,  \qquad     \Delta E^{\rm coll}_{q,g}({\rm tot.})   = \int_0^{\omega_{\rm min}} d\omega  \int_0^{R_{\rm max} }  dr \;     \omega  \frac{dN^g_{q,g}(\omega,r)}{d\omega dr}   .
\label{collshower}
\end{equation} 
The collisional energy loss that  Eq.~(\ref{collshower}) refers to is the one of the full medium induced parton shower.  From the perspective of reconstructed jets, however, only the collisional energy loss of the medium-induced parton shower that falls inside the jet cone of radius $R$ will modify the 
observed cross sections. Thus, when collisional energy losses are included   the  out-of-cone energy fraction of the medium-induced shower is
\begin{equation}
f_{q,g}^{\rm loss}(R;{\rm rad+coll}) = 1-  \left(  \int_0^{R }  dr  \int_{\omega_{\rm min}}^E  d\omega \,     \frac{dN^g_{q,g}(\omega,r)}{d\omega dr}    \right)  \Bigg/
 \left(  \int_0^{R_{\rm max} }   dr  \int_0^E  d\omega \,     \frac{dN^g_{q,g}(\omega,r)}{d\omega dr}   \right) \;.
\label{fradcol}
\end{equation} 
Clearly, the expression above reduces to  Eq.~(\ref{frad}) when $\omega_{\rm min} = 0$.  This concludes the discussion of 
Eq.~(\ref{eq:modify}).

\section{Phenomenological results}
\label{numerical}
In this section we present our phenomenological results and provide detailed comparison with the most recent experimental measurements for the isolated $\gamma$-tagged and $Z^0$ boson-tagged jet production in Pb+Pb collisions at $\sqrt{s_{NN}}=5.02$ TeV at the LHC.

In the absence of in-medium interactions one expects, to leading order in perturbative QCD,
that the transverse momentum of the vector boson is balanced by the transverse momentum of the jet,
$p_{T}^V = p_{T}^J$. Next-to-leading order processes, and  the development of parton showers in general,
break this equality. Jet reconstruction algorithms, jet radius reconstruction choice, experimental cuts,  
and detector resolution effects can all affect the exact differential distribution of 
$d\sigma / d p_{T}^V dp_{T}^J$. Still, the downshift of this distribution to smaller values of 
$p_{T}^J$ in general or the downshift of the peak in ${\rm x}_{\rm JV}=p_T^J/p_T^V$ space  are currently the 
best proxies for jet energy loss. This so-called transverse momentum imbalance $\rm x_{JV}$ distribution can be obtained from the double differential distribution of V+jet cross section 
\bea
\frac{d\sigma}{d{\rm x_{JV}}} =& \int_{p_{T}^{J, \rm min}}^{p_{T}^{J, \rm max}} d p_{T}^J
\frac{p_T^J}{\rm{x}_{JV}^2}   \frac{d\sigma(p_T^V=p_T^J/{\rm x_{JV}}, p_T^J)}{dp_T^V dp_T^J} \;,
\label{sigAA}
\eea
where $p_{T}^{J, \rm min}$ and $p_{T}^{J, \rm max}$ are matched to the desired cuts of the  experimental measurements. 

In Fig. \ref{fig:XJZ} we plot the normalized momentum imbalance distributions for the $Z^0$+jet channel (normalized by the $Z^0$ boson cross section) in both p+p and Pb+Pb collisions at the LHC, and compare the calculations to the CMS measurements~\cite{Sirunyan:2017jic}. Here, the black dashed histogram shows the Pythia 8 simulation for the p+p baseline, and the black solid points represent the CMS results. One can see that the $\rm x_{JZ}$ distribution from Pythia 8 simulation is narrower than the one measured by the CMS experiment for the p+p reference. We anticipate that this is mainly due to detector resolution effects that  have not been unfolded in the data analysis  \footnote{By applying the same smearing functions, as those that experiments apply to Monte Carlo simulations,  to our calculated 3-D $p_T$ distributions for p+p and Pb+Pb collisions, we expect to get broader ${\rm x_{JZ}}$ distributions which would bring the curves for both p+p and Pb+Pb closer to the data points.}.
The results of our theoretical calculations in Pb+Pb collisions are shown in green and magenta histograms, which correspond to jet-medium coupling strengths $g=2.0$ and $g=2.2$, respectively. These values have worked well in describing the single inclusive hadron \cite{Chien:2015vja,Kang:2014xsa}, heavy flavor mesons~\cite{Kang:2016ofv}, and jet suppression data~\cite{Kang:2017frl} at the LHC. In the implementation of energy loss effects, we have included both medium-induced radiative energy loss and energy dissipation of parton showers through collisional interactions between the jet and the medium, detailed description of these two energy loss effects can be found in the last section. By comparing Pb+Pb to p+p results, one can clearly see the downshift of ${\rm x_{JV}}$, 
as shown in Fig.~\ref{fig:XJZ}, which agrees with the data quantitatively in terms of the difference between p+p and Pb+Pb. This downshift can be easily explained by the nature of energy loss effects. The $Z^0$-boson escapes out of the medium unscathed, while part of the energy of away-side parton shower is redistributed outside of the jet cone. This reduces the jet transverse momentum and  results in the downshift of the ${\rm x_{JV}}$ distribution in Pb+Pb collisions.  
\begin{figure}[!t]
\psfig{file=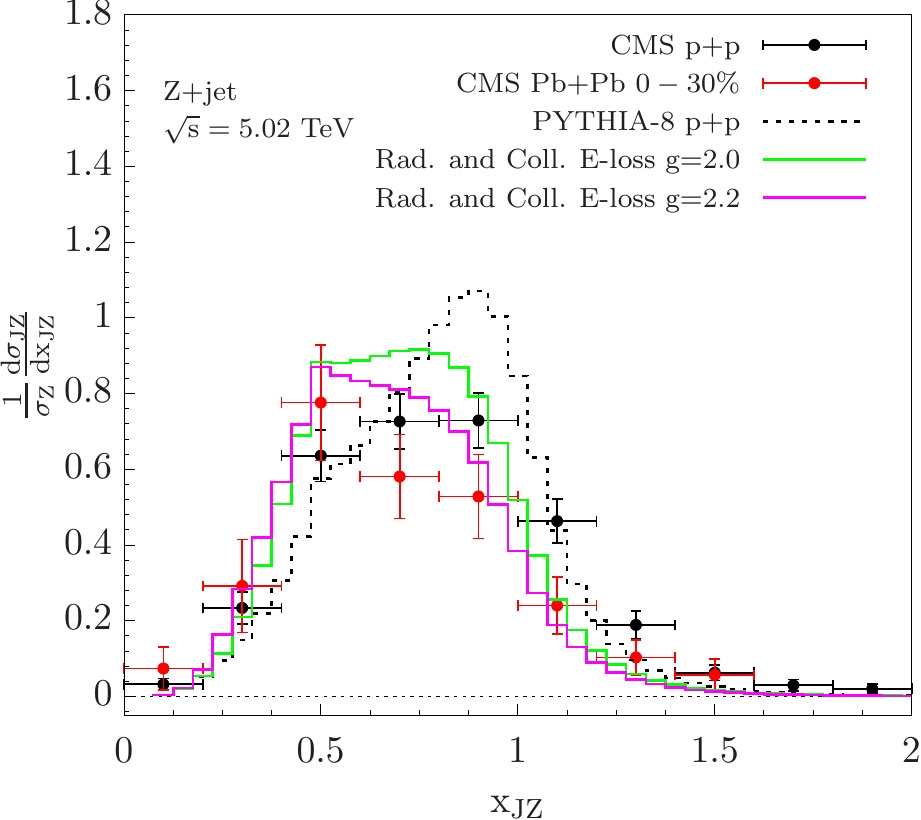, width=2.5in}
\caption{The $Z^0$-tagged jet asymmetry distribution at $\sqrt{s}=5.02$ TeV in p+p (black) and Pb+Pb (red) collisions at the LHC. The jet radius parameter is $R=0.3$, and the transverse momenta of the $Z^0$ boson and the jet are $p_T^Z>60$~GeV and 
$p_T^J>30$~GeV, respectively. The p+p baseline is simulated by Pythia 8 and shown by the black dashed line. The theoretical results for Pb+Pb collisions with two different jet-medium coupling strength are shown by the green ($g=2.0$) and magenta ($g=2.2$) histograms. The data is from the CMS collaboration~\cite{Sirunyan:2017jic}.}
\label{fig:XJZ}
\eef

To further quantify the downshift of the ${\rm x_{JV}}$ distribution, we define the mean value of ${\rm x_{JV}}$,
\bea
\langle {\rm x_{JV}} \rangle = \left.\left(\int d{\rm x_{JV}} {\rm x_{JV}} \frac{d\sigma}{d{\rm x_{JV}}}\right)\right/\left(\int d{\rm x_{JV}} \frac{d\sigma}{d{\rm x_{JV}}}\right).
\eea
In Table~\ref{table-1} we show the difference for $\langle {\rm x_{JV}} \rangle$ in p+p and Pb+Pb collisions, i.e., 
\bea
\Delta\langle {\rm x_{JV}}\rangle = \langle {\rm x_{JV}}\rangle_{\rm pp} -\langle {\rm x_{JV}}\rangle_{\rm PbPb}. 
\eea
The positive values of  $\Delta \langle {\rm x_{JV}} \rangle$ represent  downshifts of the ${\rm x_{JV}}$ distribution, and they are consistent with the experimental data within the measurement uncertainties for different $p_T^Z$ cuts. From our theoretical results, we can see the $p_T^Z$ cut dependence of  $\Delta \langle {\rm x_{JV}}\rangle$, it gets larger with the increase of $p_T^Z$ cut. However, this can't be clearly identified within the current experimental error bars. 

\begin{table}
\caption{Theoretical results for the difference of the average ${\rm x_{JZ}}$ between p+p and Pb+Pb central collisions ($0-30\%$). The center of mass energy is $\sqrt{s} = 5.02$ TeV, the transverse momentum cut for the recoil jet is $p_T^J>30$ GeV. }
\begin{center}
\begin{tabular}{l  | c | c | c | c }
\hline
\hline
\multirow{4}{*}{}
&\multicolumn{4}{ c }{$\Delta\langle{\rm x_{JZ}} \rangle $}  \\
\hline
$~p_T^Z$ (GeV)& ~$40 - 50$ ~ & ~ $50 - 60$~
  & ~ $60 - 80$ ~ & ~ $80 - 120$ ~ \\ 
 \hline
CMS \cite{Sirunyan:2017jic} & 0.061$\pm$0.059 & 0.123$\pm$0.051  & 0.124$\pm$0.052 & 0.068$\pm$0.042 \\
Rad. + Coll. $g = 2.0$ & 0.022 & 0.050 & 0.075 & 0.086 \\
Rad. + Coll. $g = 2.2$ & 0.024 & 0.058 & 0.093 & 0.119 \\
\hline
\hline
\end{tabular}
\end{center}
\label{table-1}
\end{table}

\begin{figure}[!t]
\psfig{file=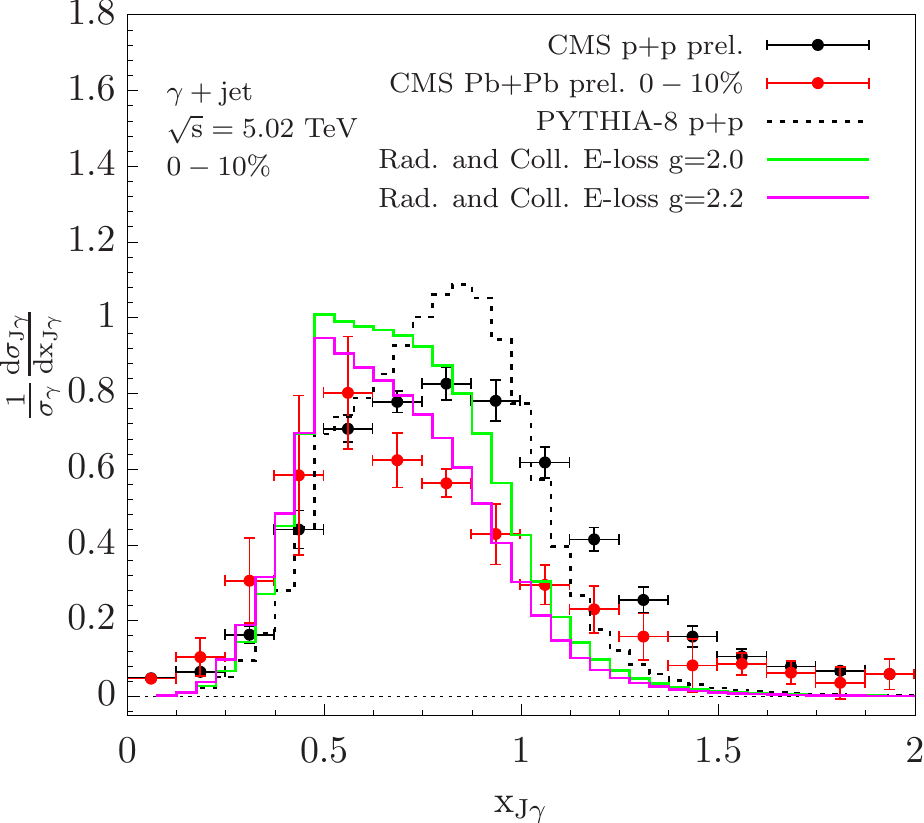, width=2.5in}
\hskip 0.2in
\psfig{file=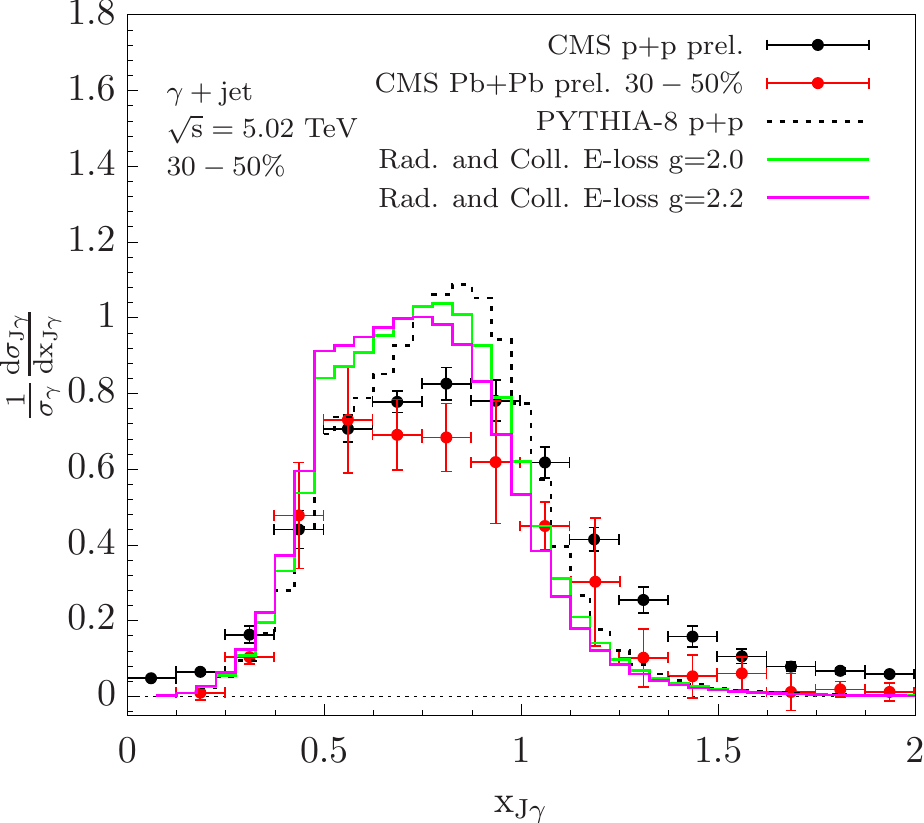, width=2.5in}
\caption{The isolated photon-tagged jet asymmetry distributions are shown and compared to CMS data in central (left) and semi-central (right) collisions~\cite{CMS:2016ynj}. The transverse momenta for the isolated photon and the jet are $p_T^{\gamma}>60$~GeV and $p_T^J>30$~GeV, respectively. The jet radius parameter is $R=0.3$. The p+p baseline, simulated by Pythia 8, is shown in the black dashed line. The theoretical results for Pb+Pb collisions with two different jet-medium coupling strengths are shown by green ($g=2.0$) and magenta ($g=2.2$) lines. }
\label{fig:xjphoton_cms}
\eef

\begin{figure}[!t]
\psfig{file=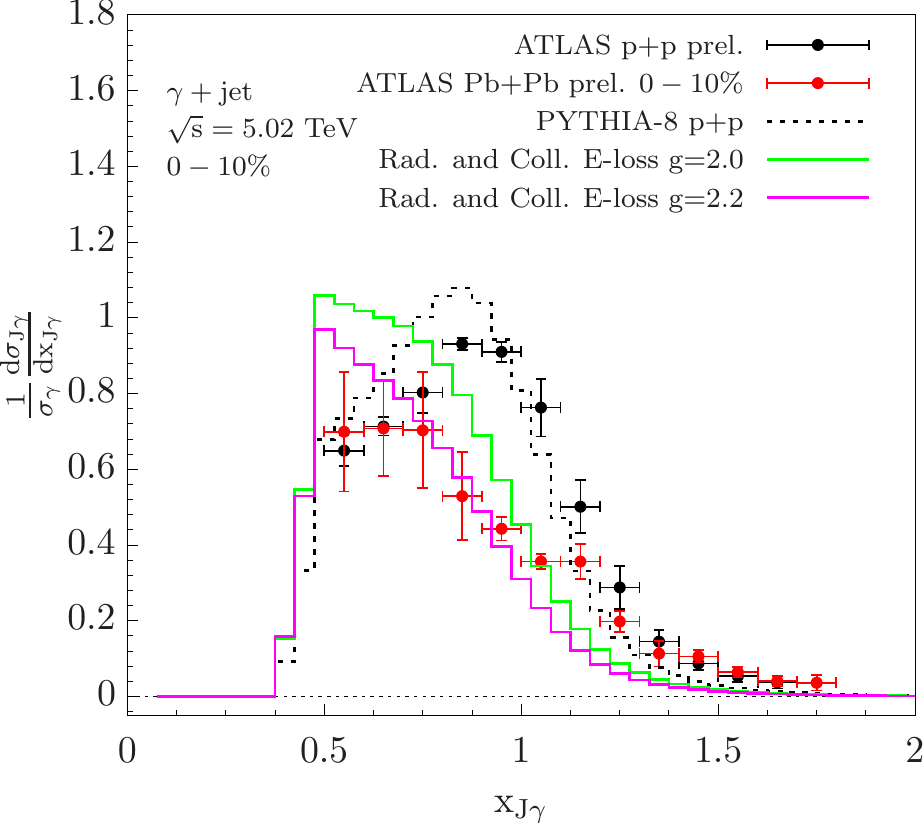, width=2.5in}
\hskip 0.2in
\psfig{file=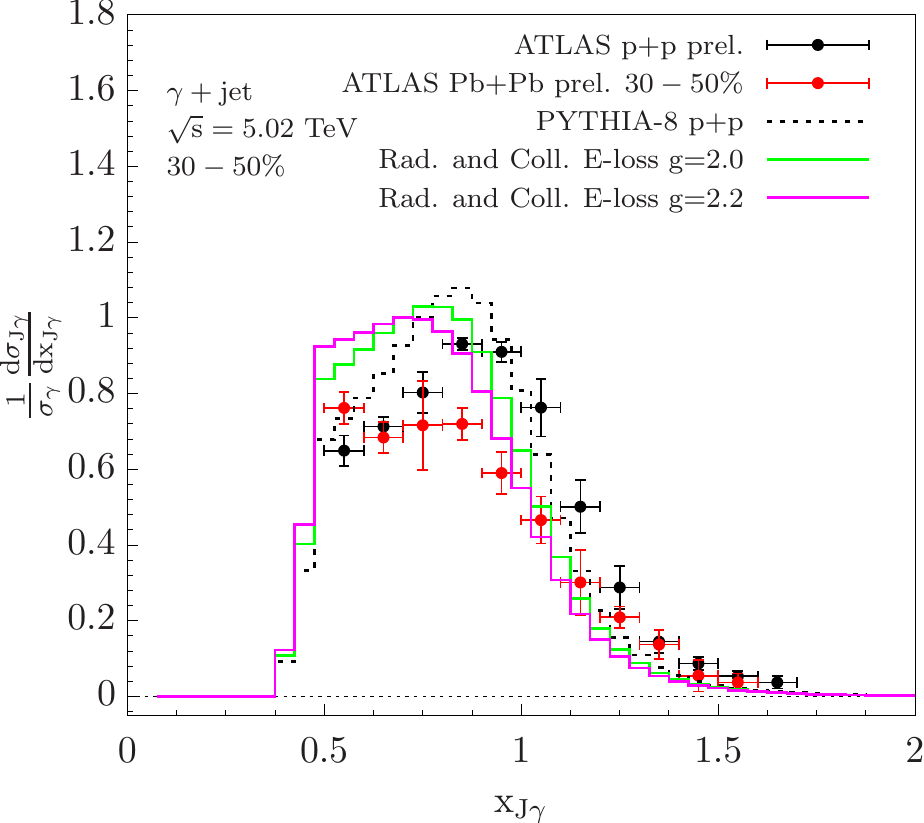, width=2.5in}
\caption{Same as in Fig. \ref{fig:xjphoton_cms}, but for comparison to ATLAS data with jet radius $R=0.4$~\cite{ATLAS:2016tor}.}
\label{fig:xjphoton_atlas}
\eef

We also evaluated the cross section for isolated-$\gamma$-tagged jet production in p+p and Pb+Pb collisions at the LHC. The comparisons to CMS and ATLAS measurements are shown in Figs.~\ref{fig:xjphoton_cms} and \ref{fig:xjphoton_atlas}, respectively. Notice that the recoil jet is reconstructed with the anti-$k_T$ algorithm with $R=0.4$ (0.3) at ATLAS (CMS). We have considered both central collisions ($0-10\%$) and semi-central ($30-50\%$) collisions. In general, theoretical calculations of the difference between   
$ {\rm x_{J\gamma}}$ distributions in p+p and Pb+Pb are quite compatible with what is seen in experimental data.  One exception is  that we didn't see as significant nuclear modifications in semi-central ($30-50\%$) collisions  as  present in the ATLAS measurements in Fig.~\ref{fig:xjphoton_atlas} (right).  We have also computed $\Delta\langle {\rm x_{J\gamma}}\rangle$ and the numerical values are given in Table~\ref{table-2} for different cuts on $p_T^{\gamma}$. We see similar behavior in the ${\rm x_{J\gamma}}$ distribution for isolated $\gamma$+jet production as  in the ${\rm x_{JZ}}$ distribution for $Z^0$+jet production. This is expected, as both processes are dominated by Compton scattering, which leads to similar energy loss effects.

\begin{table}
\caption{Theoretical results for the difference of averaged ${\rm x_{J\gamma}}$ between p+p and Pb+Pb central collisions ($0-30\%$). The center-of-mass energy is $\sqrt{s} = 5.02$ TeV, the transverse momentum cuts for the recoil jet is $p_T^J>30$ GeV. }
\begin{center}
\begin{tabular}{l  | c | c | c | c | c }
\hline
\hline
\multirow{4}{*}{}
&\multicolumn{5}{ c }{$\Delta\langle {\rm x_{J\gamma}} \rangle $}  \\
\hline
$~p_T^{\gamma}$ (GeV)& ~$40 - 50$ ~ & ~ $50 - 60$~
  & ~ $60 - 80$ ~ & ~ $80 - 100$ ~ & ~ $100 - 120$ ~ \\ 
 \hline
CMS prel. \cite{CMS:2016ynj} & 0.008$\pm$0.074 & 0.043$\pm$0.069  & 0.081$\pm$0.059 & 0.054$\pm$0.044 & 0.115$\pm$0.047 \\
Rad. + Coll. $g = 2.0$ & 0.021 & 0.044 & 0.065 & 0.075 & 0.065 \\
Rad. + Coll. $g = 2.2$ & 0.025 & 0.055 & 0.085 & 0.103 & 0.115 \\
\hline
\hline
\end{tabular}
\end{center}
\label{table-2}
\end{table}

\begin{figure}[!t]
\psfig{file=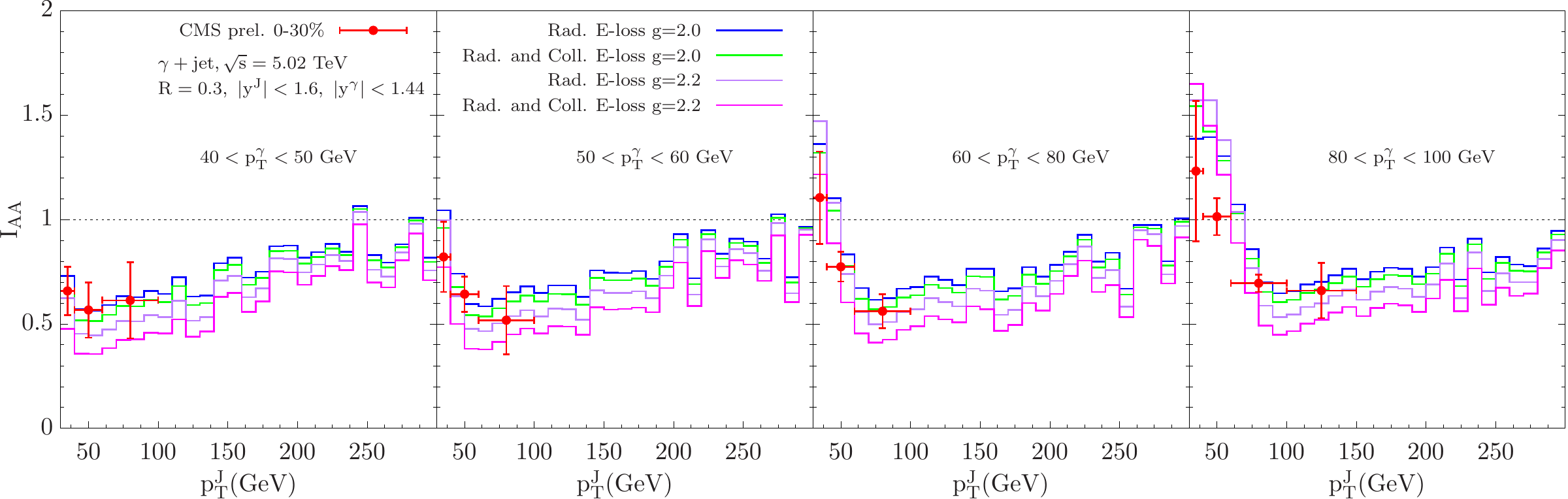, width=7.0in}
\caption{The transverse momentum cuts dependence of ${\rm I_{AA}}$ are shown comparing to CMS data. We have chosen four different setups of energy loss effects: with and without collisional energy loss for both $g=2.0$ and $g=2.2$.}
\label{fig:IAA-photonjet}
\eef

Another classical observable to quantify nuclear modification effects in V+jet systems is ${\rm I_{AA}}$, which is defined as ratio of the tagged differential cross section in A+A collisions to the binary collision scaled p+p result,
\bea
{\rm I_{AA}} = \left.\frac{1}{\langle N_{\rm bin} \rangle}\frac{d\sigma^{AA}}{[p_T^V]dp_T^J} \right/ \frac{d\sigma^{pp}}{[p_T^V]dp_T^J}\, ,
\eea
where $\langle N_{\rm bin} \rangle$ is the average number of binary nucleon-nucleon collisions for a given centrality. In this notation
we imply that the transverse momentum of the vector boson is integrated in the appropriate range
 \bea
 \frac{d\sigma}{[p_T^V]dp_T^J} \equiv   \int_{p_T^{V,min}}^{p_T^{V,max}} \frac{d\sigma}{dp_T^V dp_T^J} \;.
\eea
Our theoretical calculations for ${\rm I_{AA}}$ in isolated $\gamma$+jet production in $0-30\%$ Pb+Pb collisions are shown in Fig.~\ref{fig:IAA-photonjet}, and compared to CMS experimental data. We find that our results agree with data for a wide kinematic range. In each $p_T^{\gamma}$ window, the energy loss effects are shown in four curves with different colors, which correspond to a combination of two different jet-medium coupling strength, $g=2.0$ and $g=2.2$, as well as the situations where we either include or exclude the collisional energy loss effects in our calculations. As one has expected, the energy loss effect is more pronounced when we include collisional energy loss and a larger jet-medium coupling strength. One can see clearly in Fig. \ref{fig:IAA-photonjet} that there is a sensitive kinematical dependence of ${\rm I_{AA}}$. The largest suppression is observed along the diagonal region of the transverse momenta of the trigger $\gamma$ and the recoil jet: $p_T^{\gamma} \approx p_T^{J}$. This arises from the steeper falling cross section in the transverse momenta diagonal region. As we expect, the cross section in the region $p_T^J>p_T^{\gamma}$ is suppressed, and enhanced in $p_T^J<p_T^{\gamma}$. This is characteristic of in-medium tagged-jet dynamics. 
We further present theoretical predictions on the nuclear modification factor ${\rm I_{AA}}$ for $Z^0$+jet in Fig. \ref{fig:IAA-zjet}, which show similar $p_T^{Z}$ and $p_T^{J}$ dependence as those observed in $\gamma$+jet process. 
\begin{figure}[!t]
\psfig{file=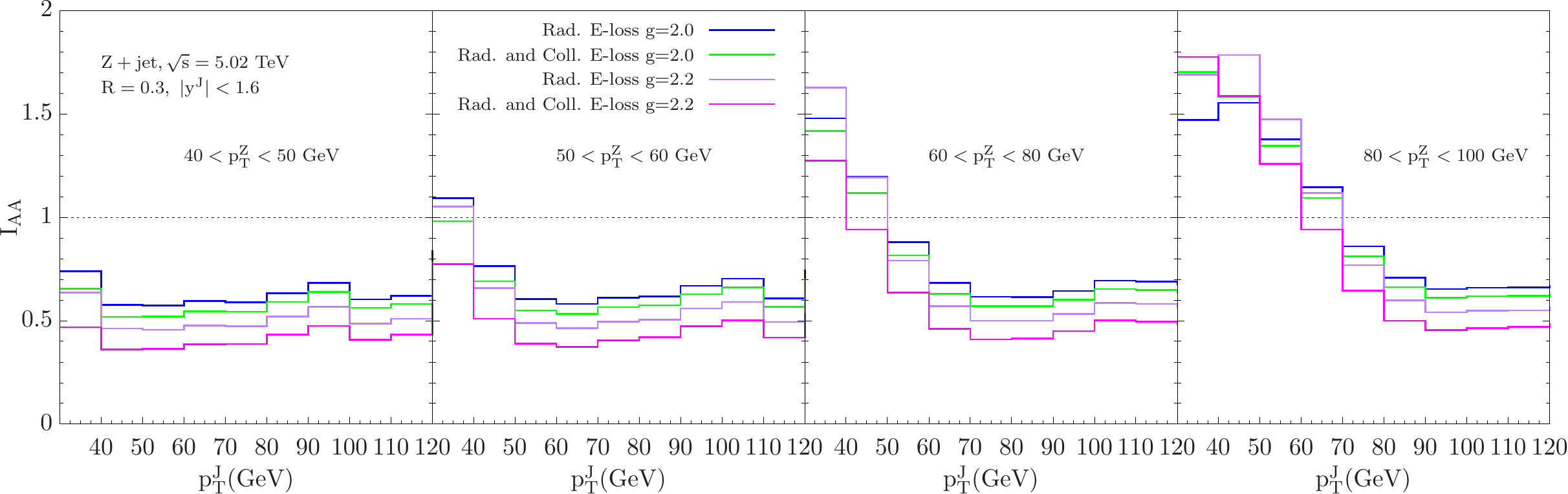, width=7.0in}
\caption{The predicted transverse momentum cuts dependence of ${\rm I_{AA}}$ for $Z^0$+jet in central ($0-30\%$) Pb+Pb collisions at $\sqrt{s_{NN}}=5.02$ TeV. The jet radius parameter is $R=0.3$. }
\label{fig:IAA-zjet}
\eef


Taking into account the observables that we have investigated, the ${\rm x_{JV}}$ momentum imbalance distributions, the 
mean  ${\rm x_{JV}}$ shift, and the tagged jet modification  ${\rm I_{AA}}$  we find that data favors coupling strengths between
the jet and the medium in the range $g = 2.0$ to $ g=  2.2$ (corresponding to $\alpha_s = 0.32 $ to  $\alpha_s = 0.39 $ at tree level). 
While the asymmetry distributions prefer the larger values of the coupling strength $g$, the  ${\rm I_{AA}}$ distributions prefer
smaller values of $g$.  Due to the complexity of the physics involved in heavy ion collisions, every theoretical calculation is
bound to have model dependence. However, the amount of out-of-cone energy redistribution due to radiative and collisional 
processes needed for modification comparable to experimental measurements is relatively robust since it only depends on the 
differential transverse momentum distribution of the recoiling jet and the proper inclusion of the Jacobian factor that accounts 
for the energy loss in Eq.~(\ref{eq:modify}). 
We present  in Table~\ref{table-3}  the results for the mean out-of-cone energy loss of  prompt quark-initiated and prompt gluon-initiated 
100~GeV jets  of small radius $R=0.3$  in central 0-10\%
Pb+Pb collisions at  $\sqrt{s_{NN}}=5.02$~TeV.   One  caveat that we must point out is  that these numbers represent the upper limits. 
The reason for that is that  multi-gluon fluctuations lead to effective 
energy losses smaller than the mean.  We find that radiative energy losses dominate, however collisional energy loss can be 
as large as 40\% correction to the radiative energy loss. This effect arises from the  high gluon multiplicity in the medium-induced 
parton shower, which amplifies collisional energy losses. This can be clearly seen by comparing the two different couplings $g$ 
between the jet and the medium. The fractional growth of the out-of-cone radiation when we include collisional energy loss is larger 
for $g=2.2$ in comparison to $g=2.0$ because the multiplicity of the medium-induced parton shower is larger in addition to the 
collisional energy loss for each individual gluon being larger itself. Last but not least, by comparing the magnitudes of  out-of-cone
energy loss for different scenarios we see that the strength of nuclear modification of the observables indeed follows the ordering of
$ \langle  \Delta E_{q,g}^{\rm out} \rangle$.
 
\begin{table}
\caption{The mean energy loss for 100~GeV jets of $R=0.3$ initiated by prompt quarks and gluons rounded to the nearest GeV. Central $0-10\%$ Pb+Pb collisions at  $\sqrt{s_{NN}}=5.02$ TeV are considered. }
\begin{center}
\begin{tabular}{l  | c | c | c | c }
\hline
\hline
\multirow{3}{*}{}
&\multicolumn{4}{ c }   {$ \langle  \Delta E_{q,g}^{\rm out} \rangle   =  E_{q,g}^{\rm jet}  \langle \epsilon \rangle    f_{q,g}^{\rm loss}(R)  $}       \\[1ex]
\hline
~~~~~~Type of E-loss ~& Rad.  g=2.0 ~ & ~ Rad.+Col.  g=2.0~ 
  & ~ Rad.  g=2.2~ &  ~ Rad.+Col.  g=2.2 ~ \\ 
 \hline
Prompt quark-initiated jet~~ & 7 GeV & 8 GeV & 10 GeV & 14 GeV  \\
Prompt gluon-initiated jet~~ & 15 GeV & 18 GeV & 21 GeV & 29 GeV \\
\hline
\hline
\end{tabular}
\end{center}
\label{table-3}
\end{table}


\section{Conclusions}
\label{summary}
In summary, in this paper we presented a new study of vector boson-tagged (either isolated $\gamma$ or $Z^0$)  jet production in Pb+Pb 
collisions at a center-of-mass energy per nucleon pair of 5.02~TeV. This work is timely since new experimental results on these final states from the LHC 
experiments are becoming available.  Within the traditional energy loss approach, by including both collisional and radiative energy loss effects, we evaluated several experimentally relevant observables: the so-called transverse momentum imbalance ${\rm x_{JV}}$ distribution modification in going 
from  p+p to Pb+Pb collisions, the related mean momentum imbalance shift  $ \Delta \langle {\rm x_{JV}} \rangle$,  and the tagged jet nuclear modification 
factor ${\rm I_{AA}}$. While some tension remains between the baseline Pythia simulations and the experimental measurements, which  at present 
are not unfolded for detector resolution effects, we found good agreement between the theoretical simulations of the modification of these  
observables for coupling strengths between the jet and the medium $g=2.0$ to $g=2.2$ and the experimental results.  This agreement is 
encouraging, and supports the emerging picture of the in-medium parton shower formation as encoded in these calculations. Both  
$\gamma$-tagged and $Z^0$-tagged jets are very effective in selecting prompt quark-initiated jets and can provide valuable information on the 
flavor dependence of parton energy loss.  We further found that while for small radius jets radiative energy loss gives the dominant contribution, 
collisional energy loss may play a significant role, especially for larger coupling strengths of the interaction between the jet and the medium.
We conclude by emphasizing that the substructure modification of $\gamma$-tagged and $Z^0$-tagged jets can differ quite substantially
from the substructure modification of inclusive jets and future experimental measurements of such observables can add significantly to our understanding
 of in-medium QCD dynamics.

\section*{Acknowledgments}
We thank Yen-Jie Lee, Qiuguang Liu, Dennis V. Perepelitsa and Kaya Tatar for helpful discussions and suggestions. This work is supported by the U.S. Department of Energy under Contract Nos.~DE-AC52-06NA25396 (Z.K., I.V.), DE-FG02-91ER40684 (H.X.), and DE-AC02-06CH11357 (H.X.).


\end{document}